\documentclass[preprint,prc,showpacs,preprintnumbers,amsmath,amssymb,floatfix]{revtex4-2}

\usepackage{amsmath,amssymb,bm}
\usepackage{graphicx}
\usepackage{xcolor}

\begin{document}
	
	\title{Vortex Nucleons as Partial-Wave Filters in Nucleon--Nucleon Scattering}
	
	\author{Jinniu Hu}
	\affiliation{School of Physics, Nankai University, Tianjin 300071, China}
	\affiliation{Shenzhen Research Institute of Nankai University, Shenzhen 518083, China}
	
	\author{Ying Zhang}
	\email{yzhang@tju.edu.cn}
	\affiliation{Department of Physics, School of Science, Tianjin University, Tianjin 300072, China}
	
	\author{Hong Shen}
	\email{songtc@nankai.edu.cn}
	\affiliation{School of Physics, Nankai University, Tianjin 300071,  China}
	
	\begin{abstract}
		We propose vortex nucleon scattering as an angular-momentum-resolved probe of nucleon--nucleon partial waves. Using the standard $LSJ$ partial-wave $S$ matrix as input, we show that an on-axis vortex incident state with a fixed orbital angular-momentum projection $m_L=\ell$ imposes the direct selection rule $L\geq |\ell|$ on the initial nucleon--nucleon partial waves. As a result, the initial $S$ wave is excluded for $\ell=1$, while both the initial $S$ and $P$ waves are excluded for $\ell=2$. The underlying phase shifts are not modified. Instead, the vortex external state changes how the ordinary partial waves are projected into the scattering amplitude. We further analyze off-axis scattering, where the displacement of the target from the vortex axis introduces Bessel-function weights and partially relaxes the on-axis selection rule. These results suggest that vortex nucleons can provide a new experimental handle on the partial-wave content of the strong nucleon--nucleon interaction.
	\end{abstract}
	
	\maketitle
	
	Nucleon--nucleon ($NN$) scattering is one of the most direct probes of the strong interaction at low and intermediate energies. Modern descriptions are based on high-precision phase-shift analyses and realistic interactions, including the Nijmegen potential, the Argonne interaction, the charge-dependent Bonn (CD-Bonn) potential, and chiral effective field theory based on the low-energy QCD theory \cite{Machleidt1989AdvNuclPhys,Stoks1993PRC,Wiringa1995PRC,Machleidt2001PRC,Epelbaum2009RMP,MachleidtEntem2011PhysRep}. In these approaches, the $NN$ amplitude is naturally decomposed into $LSJ$ partial waves, and plane-wave observables are obtained from a coherent sum over all allowed channels. This success also makes a basic question natural. Can one use a structured incident state to select partial-wave components more directly?
	
	Vortex, or twisted, particles provide such an external angular-momentum scheme. A vortex state carries an azimuthal phase factor
	\begin{equation}
		\psi(\rho,\phi,z)\propto e^{i\ell\phi},
		\label{eq:vortex_phase}
	\end{equation}
	and hence a well-defined orbital angular momentum (OAM) projection along its propagation axis. Vortex electron beams are the best developed matter-wave realization. They have been generated experimentally and used or proposed in electron microscopy, magnetic and chiral probes, phase-sensitive scattering, and atomic collision studies \cite{Verbeeck2010Nature,Bliokh2017PhysRep,Lloyd2017RMP,Bliokh2011PRL,VanBoxem2014PRA,Ivanov2012PRD,Ivanov2023PRA,Tolstikhin2019PRA,Strnat2024arXiv,Strnat2025Atoms,Harris2025JPB}. These studies indicate that the transverse momentum and OAM of a structured beam can expose information that is averaged out in plane-wave pictures.
	
	Vortex and angular-momentum-resolved probes are also entering nuclear physics. The OAM of neutrons has been controlled experimentally \cite{Clark2015Nature}. Twisted-neutron and twisted-photon reactions have been discussed for radiative capture, photodisintegration, Schwinger scattering, and elastic neutron-nucleus scattering, where the vortex structure can affect cross sections, spin asymmetries, polarization, and phase-sensitive observables \cite{Afanasev2018JPG,Afanasev2019PRC,Afanasev2021PRC}. Related ideas have been proposed for angular-momentum-resolved inelastic electron scattering from nuclear giant resonances, photon-vortex generation, and twisted-light excitation of the $^{229}$Th nuclear clock transition \cite{Lu2025PRL,Maruyama2023PRResearch,Kirschbaum2024PRC}. Existing work, however, mainly concerns electron-atom, electron-nucleus, neutron-nucleus, or photon-nucleus reactions.
	
	Here we ask how a vortex external state acts on the two-nucleon problem. In this work, we use the ordinary $NN$ partial-wave $S$ matrix as the dynamical input and replace only the incident external state by a Bessel-type vortex nucleon, while the interaction is not modified. We first construct the fixed-OAM vortex amplitude from the $LSJ$ expansion and then derive the selection rule, making clear that it follows from the external-state projection rather than from a new assumption about the $NN$ interaction. We then discuss off-axis scattering, where a target displacement introduces Bessel-function weights and relaxes the on-axis rule.
	
	We consider elastic two-nucleon scattering in the center-of-mass frame. The relative momenta are denoted by $\mathbf q_i$ and $\mathbf q_f$, and the two-nucleon spin state is written in the coupled basis $|S\mu_S\rangle$. The ordinary $NN$ dynamics is specified by the partial-wave matrix $S^J_{L'S',LS}$. It is convenient to introduce
	\begin{equation}
		\begin{aligned}
			K^J_{L'S',LS}
			=&\,S^J_{L'S',LS}
			-\delta_{L'L}\delta_{S'S} .
		\end{aligned}
		\label{eq:K_matrix}
	\end{equation}
	Here $K^J_{L'S',LS}$ is the scattering part of the partial-wave $S$ matrix. In the numerical calculations below, $S^J_{L'S',LS}$ is constructed from the PWA93 phase shifts and mixing angles tabulated by NN-Online, which are based on the Nijmegen partial-wave analysis \cite{Stoks1993PRC,NNOnline}. The plane-wave and vortex calculations use the same on-shell phase shifts. The differences discussed below therefore arise solely from the angular-momentum projection of the external state.
	
	The fixed-OAM incident state is taken as a Bessel-type vortex state,
	\begin{equation}
		|\kappa,q_z,\ell\rangle
		=
		\int_0^{2\pi}\frac{d\phi_i}{2\pi}
		e^{i\ell\phi_i}
		|\mathbf q_i(\phi_i)\rangle,
		\label{eq:bessel_state}
	\end{equation}
	where $|\mathbf q_i(\phi_i)\rangle$ denotes a plane-wave state with momentum
	\begin{equation}
		\begin{aligned}
			\mathbf q_i(\phi_i)
			=&\,q(\sin\theta_\kappa\cos\phi_i,
			\sin\theta_\kappa\sin\phi_i,
			\cos\theta_\kappa),\\
			\theta_\kappa=&\,\arctan(\kappa/q_z).
		\end{aligned}
		\label{eq:cone_momentum}
	\end{equation}
	For an incident plane wave chosen along the quantization axis, the incoming orbital projection is $m_L=0$. In contrast, the on-axis Bessel vortex state is a coherent cone of plane waves with fixed OAM projection $\ell$. The azimuthal integral over the incident cone gives
	\begin{equation}
		\begin{aligned}
			&\int_0^{2\pi}\frac{d\phi_i}{2\pi}
			e^{i\ell\phi_i}
			Y^*_{Lm_L}(\theta_\kappa,\phi_i)=
			Y^*_{L\ell}(\theta_\kappa,0)\delta_{m_L\ell} .
		\end{aligned}
		\label{eq:oam_projection}
	\end{equation}
	This fixes the incoming orbital projection to $m_L=\ell$. The corresponding on-axis fixed-OAM amplitude is
	\begin{equation}
		\begin{aligned}
			&\mathcal A^{(\ell)}_{S'\mu'_S,S\mu_S}
			(\theta_f,\phi_f)
			=\frac{4\pi}{2iq}
			\sum_{JLL'} i^{L-L'}
			\\
			&\quad\times
			Y^*_{L\ell}(\theta_\kappa,0)
			\langle L\ell,S\mu_S|JM\rangle
			K^J_{L'S',LS}
			\\
			&\quad\times
			\langle L'm'_L,S'\mu'_S|JM\rangle
			Y_{L'm'_L}(\theta_f,\phi_f) .
		\end{aligned}
		\label{eq:onaxis_oam_amp}
	\end{equation}
	with
	\begin{equation}
		M=\ell+\mu_S,
		\qquad
		m'_L=M-\mu'_S .
		\label{eq:onaxis_projection}
	\end{equation}
	It is explicit from Eq.~\eqref{eq:onaxis_oam_amp} that the vortex external state replaces the plane-wave projection $m_L=0$ by $m_L=\ell$. Since the spherical harmonic and Clebsch--Gordan coefficient vanish unless $|m_L|\leq L$, the fixed projection $m_L=\ell$ requires
	\begin{equation}
		L\geq |\ell| .
		\label{eq:selection_rule}
	\end{equation}
	Thus a vortex nucleon with $\ell=1$ removes the initial $S$-wave contribution, while $\ell=2$ removes both the initial $S$- and $P$-wave contributions. This replacement leaves the underlying phase shifts unchanged but reorganizes their contribution to the physical amplitude.
	
	The product-spin scattering matrix $\mathcal M^{(\ell)}$ is constructed by recoupling the spin basis from $|S\mu_S\rangle$ to $|\sigma_1\sigma_2\rangle$. The $NN$ differential cross section is written as
	\begin{equation}
		\begin{aligned}
			\frac{d\sigma_\ell}{d\Omega_f}
			=&\,\frac14 {\rm Tr}\left[
			\mathcal M^{(\ell)}
			\mathcal M^{(\ell)\dagger}
			\right] .
		\end{aligned}
		\label{eq:cross_section}
	\end{equation}
	Spin observables follow from the same matrix by inserting the appropriate Pauli operators. Similar trace expressions are used for final polarizations and spin-correlation coefficients.
	
\begin{figure}[t]
		\centering
		\includegraphics[width=1.1\linewidth]{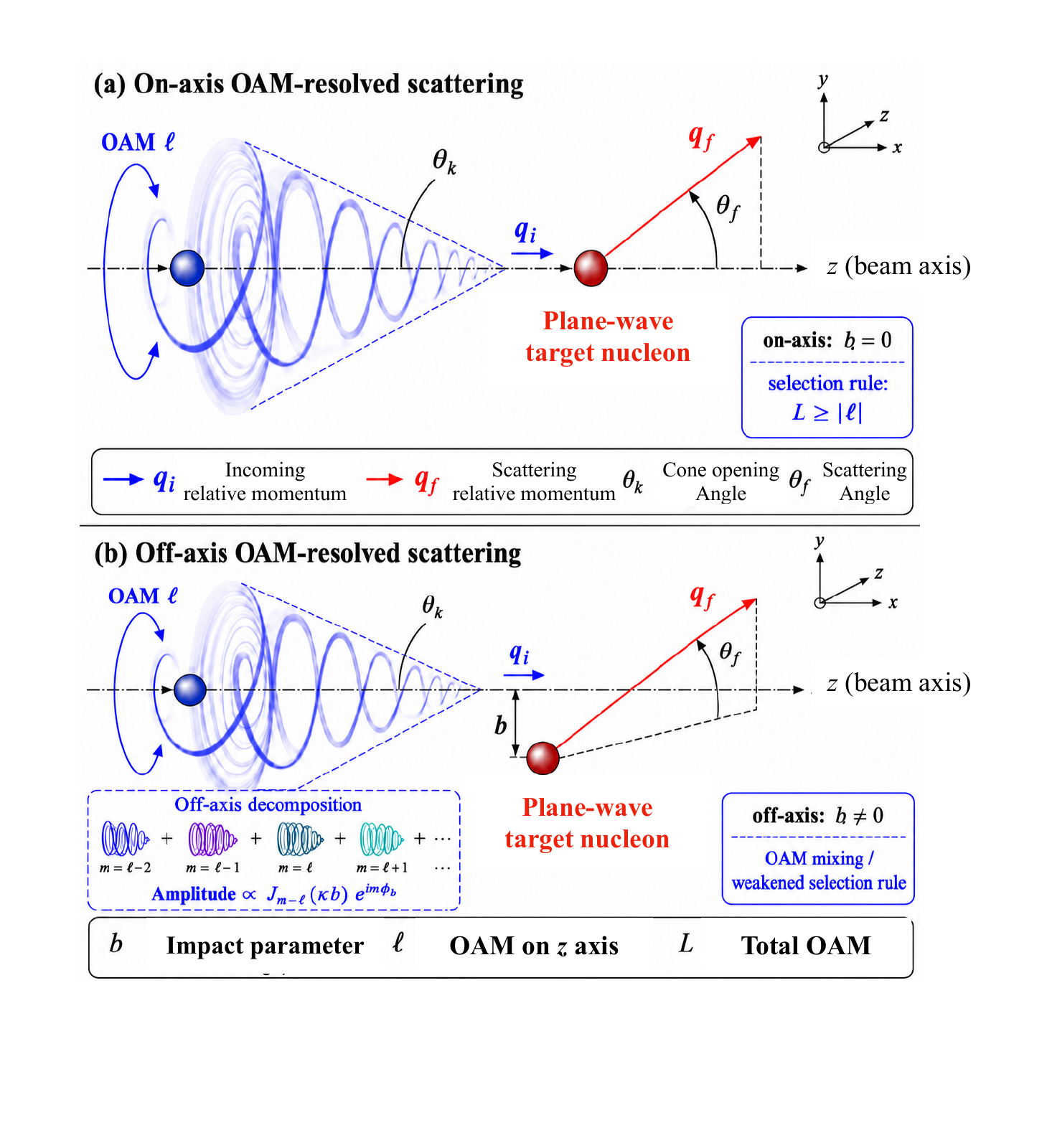}
		\caption{Schematic illustration of on-axis and off-axis OAM-resolved vortex nucleon scattering. The angle $\theta_\kappa$ is the vortex cone opening angle, while $b$ is the transverse displacement of the target from the vortex axis.}
		\label{fig:schematic}
	\end{figure}

	The vortex nucleon scattering process used in the present calculation is sketched in Fig.~\ref{fig:schematic}. The incident vortex nucleon is a coherent superposition of plane-wave components whose momenta lie on a cone with opening angle $\theta_\kappa$. The beam axis is chosen as the quantization axis. In the on-axis scattering process of Fig.~\ref{fig:schematic}(a), the target is located at the vortex center, $b=0$, and the vortex charge fixes the orbital projection sampled at the target to $m_L=\ell$. The scattered relative momentum $\mathbf q_f$ is observed at the angle $\theta_f$. This is the cleanest realization of OAM filtering.
	
	For the off-axis scattering in Fig.~\ref{fig:schematic}(b), the target is shifted from the vortex axis by the transverse distance $b$. Only the projection of this displacement in the scattering plane is drawn; its azimuthal direction is denoted by $\phi_b$ in the formulae below. The nuclear interaction remains local at the target, but the vortex state must be re-expanded about the target center. This translation mixes several target-centered orbital projections and therefore weakens the sharp on-axis selection rule.

	\begin{figure}[t]
		\centering
		\includegraphics[width=\linewidth]{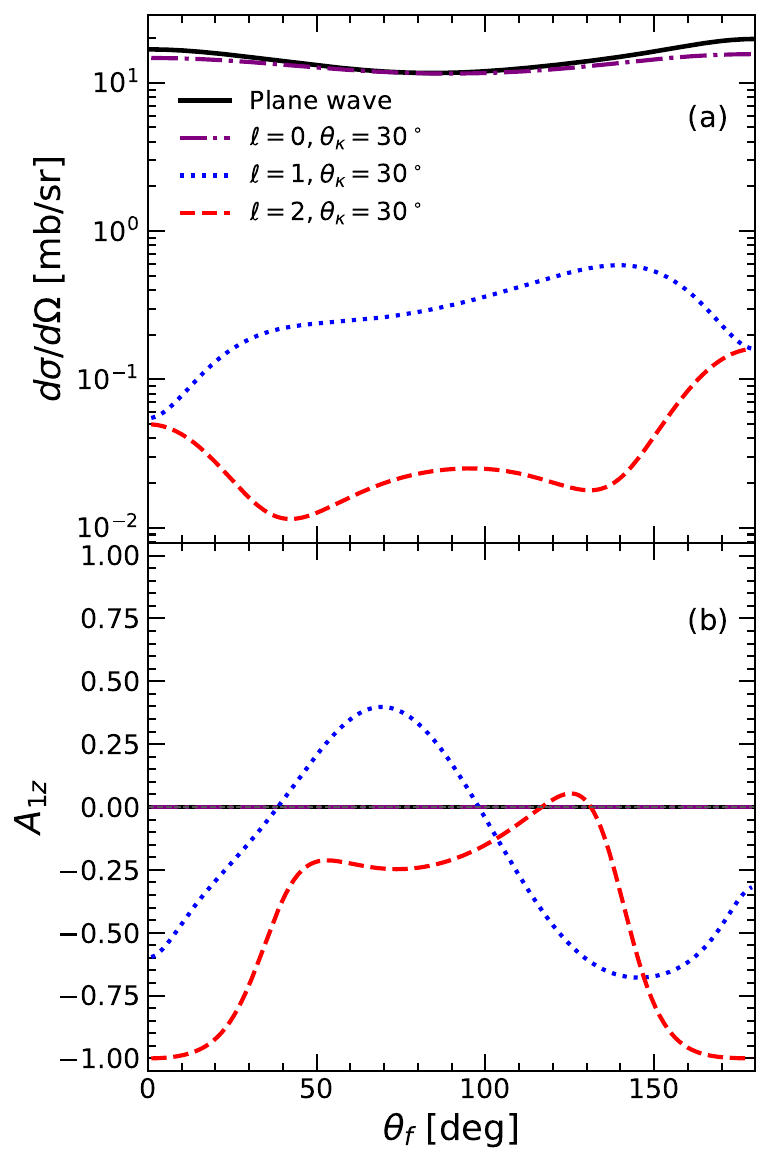}
		\caption{
			Differential cross section and longitudinal beam analyzing power \(A_{1z}\) for plane-wave and vortex \(np\) scattering at \(T_{\rm lab}=50\) MeV. The vortex results are calculated at \(\theta_\kappa=30^\circ\).}
		\label{fig:dsigma_A1z}
	\end{figure}
	
	A representative example for $np$ scattering observables at $T_{\rm lab}=50$ MeV is presented in Fig.~\ref{fig:dsigma_A1z}. The ordinary plane-wave result is compared with fixed-OAM vortex states at $\theta_\kappa=30^\circ$, which separates the finite-cone effect for $\ell=0$ from the OAM-filtering effect for $\ell\neq0$.
	
	The differential cross section is given in panel (a). The $\ell=0$ vortex curve stays close to the plane-wave result but is not identical to it, because the finite cone angle weights the partial waves by $Y_{L0}^{*}(\theta_\kappa,0)$ rather than by the forward value at $\theta_\kappa=0$. The deviation is modest, about $7.4\%$ at $\theta_f\simeq30^\circ$, $1.2\%$ near $90^\circ$, and $11\%$ near $150^\circ$. Thus the finite cone angle slightly distorts the angular shape, whereas it does not produce the strong suppression associated with nonzero OAM.
	
	For $\ell=1$ and $\ell=2$, the suppression is much stronger. At $\theta_f\simeq30^\circ$, the cross section is reduced to about $1.3\%$ and $0.1\%$ of the plane-wave value, respectively. The corresponding ratios are about $2.7\%$ and $0.2\%$ near $90^\circ$, and about $3.3\%$ and $0.25\%$ near $150^\circ$. This reduction is the numerical manifestation of Eq.~\eqref{eq:selection_rule}, since the $\ell=1$ beam removes the initial $S$ wave, while the $\ell=2$ beam also removes the initial $P$ waves.
	
	The quantity plotted in panel (b) is the longitudinal beam analyzing power, $A1z$, which should be zero for the plane wave, defined by
	\begin{equation}
		\begin{aligned}
			A_{1z}
			=&\,\frac{{\rm Tr}\left[
				\mathcal M^{(\ell)}
				(\sigma_z\otimes I)
				\mathcal M^{(\ell)\dagger}
				\right]}
			{{\rm Tr}\left[
				\mathcal M^{(\ell)}
				\mathcal M^{(\ell)\dagger}
				\right]} .
		\end{aligned}
		\label{eq:A1z}
	\end{equation}
	This normalized ratio probes the spin-projection content of the surviving amplitude rather than the absolute scattering strength.
	
	The longitudinal analyzing power $A_{1z}$ as a function of $\theta_f$ is displayed in panel (b). It is essentially zero for the plane wave and for the $\ell=0$ vortex state, as expected from the usual symmetry of plane-wave scattering and the absence of a nonzero OAM projection. For $\ell=1$, however, $A_{1z}$ reaches about $0.35$ near $\theta_f\simeq60^\circ$ and $-0.67$ near $150^\circ$. For $\ell=2$, it is already about $-0.70$ near $30^\circ$ and approaches $-1$ at very backward angles. The vortex OAM therefore changes not only the magnitude of the cross section but also the spin-projection content of the remaining amplitude.
	
	In fact, the on-axis condition is idealized. If the target is displaced from the vortex axis by a transverse impact parameter $\mathbf b=b(\cos\phi_b,\sin\phi_b,0)$, each plane-wave component in the vortex superposition acquires the translation phase $\exp[-i\kappa b\cos(\phi_i-\phi_b)]$. The fixed on-axis projection is then replaced by a coherent sum over target-centered orbital projections,
	\begin{equation}
		\begin{aligned}
			C^{(\ell)}_{Lm}(\theta_\kappa,\kappa b,\phi_b)
			=&\,Y^*_{Lm}(\theta_\kappa,0)
			(-i)^{m-\ell}
			\\
			&\times J_{m-\ell}(\kappa b)
			e^{-i(m-\ell)\phi_b},
		\end{aligned}
		\label{eq:offaxis_coeff}
	\end{equation}
	{where $J_n(x)$ denotes the Bessel function of the first kind.} The corresponding off-axis fixed-OAM amplitude reads
	\begin{equation}
		\begin{aligned}
			&\mathcal A^{(\ell)}_{S'\mu'_S,S\mu_S}
			(\theta_f,\phi_f;\mathbf b)
			\\
			&=\frac{4\pi}{2iq}
			\sum_{JLL'm} i^{L-L'}
			C^{(\ell)}_{Lm}(\theta_\kappa,\kappa b,\phi_b)
			\\
			&\quad\times
			\langle Lm,S\mu_S|JM\rangle
			K^J_{L'S',LS}
			\\
			&\quad\times
			\langle L'm'_L,S'\mu'_S|JM\rangle
			Y_{L'm'_L}(\theta_f,\phi_f) .
		\end{aligned}
		\label{eq:offaxis_oam_amp}
	\end{equation}
	Here the angular-momentum projections are
	\begin{equation}
		M=m+\mu_S,
		\qquad
		m'_L=M-\mu'_S .
		\label{eq:offaxis_projection}
	\end{equation}
	Eqs.~\eqref{eq:offaxis_coeff}--\eqref{eq:offaxis_projection} indicate that finite \(b\) only changes the target-centered angular-momentum decomposition of the incident vortex state. The underlying \(NN\) interaction, encoded in { \(K^J_{L'S',LS}\)}, remains unchanged. It changes the projection of the incident vortex state onto target-centered orbital components. For $b=0$, Eq.~\eqref{eq:offaxis_coeff} gives $J_{m-\ell}(0)=\delta_{m\ell}$ and the on-axis amplitude is recovered. For $b\neq0$, several $m$ components contribute, so the sharp selection rule in Eq.~\eqref{eq:selection_rule} is partially relaxed.
	
	This off-axis relaxation has a direct physical interpretation. In the on-axis case, the target samples a definite OAM projection about the beam axis. In the off-axis case, the same vortex state is decomposed about a shifted center. Partial waves that are forbidden at $b=0$ can then re-enter with strengths controlled by $J_{m-\ell}(\kappa b)$, and the cross section can become nonmonotonic in $b$.
	
	\begin{figure}[t]
		\centering
		\includegraphics[width=\linewidth]{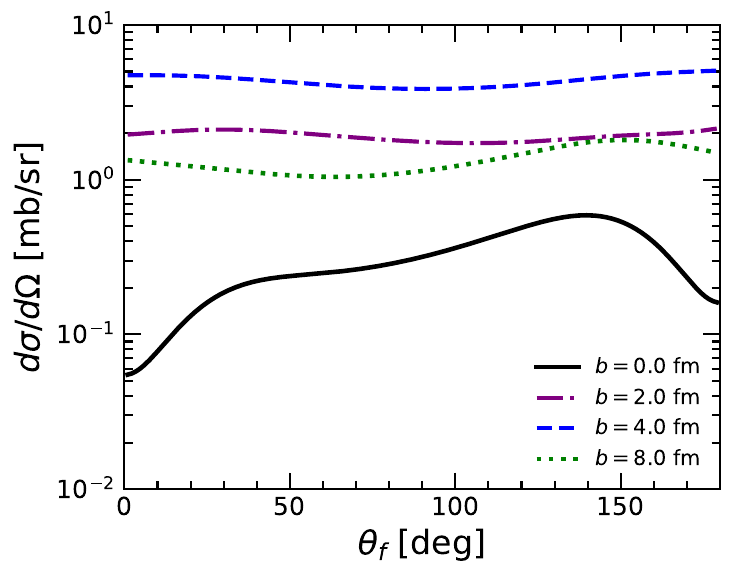}
		\caption{Off-axis differential cross sections for the fixed-OAM state with $\ell=1$ at $T_{\rm lab}=50$ MeV. The vortex cone angle is $\theta_\kappa=30^\circ$, and the target displacement direction is chosen as $\phi_b=0$.}
		\label{fig:offaxis_b}
	\end{figure}
	
	This off-axis effect is quantified in Fig.~\ref{fig:offaxis_b} for the $\ell=1$ vortex state. Here $b$ is the transverse displacement of the target from the vortex axis, and $\phi_b=0$ fixes its azimuthal direction. The on-axis result is strongly suppressed because the initial $S$ wave is excluded. It stays below about $0.6$ mb/sr over the full angular range. At finite $b$, the cross section increases by more than one order of magnitude over a broad angular region. Around $\theta_f\simeq90^\circ$, for example, it rises from roughly $0.3$ mb/sr at $b=0$ to about $1.5$--$2$ mb/sr for $b=2$ fm and about $4$ mb/sr for $b=4$ fm.
	
	This enhancement should not be interpreted as a stronger nuclear force at a larger $NN$ separation. The parameter $b$ specifies where the target samples the transverse profile of the vortex state, while it is not the interaction distance between the two nucleons. At finite $b$, the Bessel weights generate target-centered components with $m\neq\ell$. For $\ell=1$, an $m=0$ component is then allowed, and the strong low-$L$ partial waves, including the $S$ wave, can re-enter. The fact that the $b=4$ fm curve is larger than both the $b=2$ fm and $b=8$ fm curves over much of the angular range reflects the oscillatory Bessel weights rather than a monotonic distance dependence of the nuclear interaction. In a realistic finite wave packet, this fixed-$b$ response must be multiplied by the transverse beam profile and averaged over the target distribution. The observed event rate would eventually decrease when the target lies outside the transverse support of the finite vortex wave packet.
	
	Because the $NN$ amplitude is spin dependent, the OAM-induced filtering of partial waves inevitably changes the relative weights of singlet, triplet, uncoupled, and coupled spin channels. The tensor-coupled ${}^3S_1$--${}^3D_1$ channel is particularly sensitive, since its $S$-wave contribution is suppressed for an $\ell=1$ vortex state on axis, whereas a finite impact parameter can partially restore this contribution through the off-axis projection of the vortex wave function. Analyzing powers and spin-correlation observables therefore provide complementary probes of the same mechanism. As the target is displaced from the vortex axis, its overlap with the annular intensity profile increases, which enhances the reaction probability near the bright ring of the vortex beam. This behavior may be interpreted as an enhanced sensitivity to off-axis, or grazing-like, collision geometries, although the effect is intrinsically wave mechanical rather than classical.
	
	The distinction between fixed orbital angular momentum and fixed total angular momentum (TAM) deserves comment. Nucleons carry spin, so one may also construct incident vortex states with a fixed total angular-momentum projection $M_{\rm in}=m_L+\mu_S$. In such a fixed-TAM formulation, the orbital projection entering a given spin channel is $m_L=M_{\rm in}-\mu_S$, and the corresponding selection condition becomes $L\geq |M_{\rm in}-\mu_S|$. This is different from the fixed-OAM rule in Eq.~\eqref{eq:selection_rule}, because it depends explicitly on the spin projection. Thus fixed-OAM and fixed-TAM vortex nucleon states are not equivalent external-state preparations. The fixed-OAM formulation gives the most transparent partial-wave filtering mechanism, while the fixed-TAM formulation is more suitable for analyzing total-angular-momentum transfer. {See Supplemental Material for detailed derivations, additional polarization observables, and partial-wave decompositions \cite{supply}}. A systematic comparison of OAM- and TAM-resolved vortex nucleon scattering will be addressed in future work.
	
	In summary, vortex nucleons provide a new way to investigate the partial-wave structure of the strong $NN$ interaction. The standard phase shifts and mixing angles remain the dynamical input, while the vortex external state imposes an additional angular-momentum projection. For an on-axis fixed-OAM incident state, this gives the selection rule $L\geq|\ell|$, which filters the low partial waves and makes the contributions of higher-$L$ and tensor-coupled channels more visible in selected observables. Off-axis scattering introduces Bessel-function mixing and partially relaxes this rule, thereby connecting the ideal on-axis selection mechanism with more realistic beam-target geometries. The sizable changes found in both the cross section and $A_{1z}$ indicate that vortex nucleons can act not only as intensity filters but also as spin-selective probes of the $NN$ amplitude. Future work should include finite transverse wave packets, realistic target distributions, and fixed-TAM vortex states, as well as possible experimental strategies for producing and detecting vortex nucleon beams. If such beams are realized, they could provide an angular-momentum-resolved tool for isolating individual partial-wave contributions in two-nucleon scattering.
	
	This work was supported by the National Natural Science Foundation of China under Grant No. 12475149 and by the Guangdong Basic and Applied Basic Research Foundation under Grant No. 2024A1515010911.

	\bibliographystyle{apsrev4-2}
	\bibliography{references}
	
\end{document}